\documentclass{iopart}

\def\br{{\bf r}}

\begin{document}
\title[The two-dimensional Coulomb gas]{The statistical mechanics of the 
classical two-dimensional Coulomb gas is exactly 
solved\footnote{Invited talk presented at the SCCS conference 
in Santa Fe, 2002} }

\author{L. {\v S}amaj}

\address{Institute of Physics, Slovak Academy of Sciences, 
D\'ubravsk\'a cesta 9, SK-842 28 Bratislava, Slovakia}

\ead{fyzimaes@savba.sk}

\begin{abstract}
The model under consideration is a classical 2D Coulomb gas 
of pointlike positive and negative unit charges, 
interacting via a logarithmic potential.
In the whole stability range of temperatures,
the equilibrium statistical mechanics of this fluid model
is exactly solvable via an equivalence with the integrable
2D sine-Gordon field theory.
The exact solution includes the bulk thermodynamics,
special cases of the surface thermodynamics, and the large-distance 
asymptotic behavior of the two-body correlation functions.
\end{abstract}


\section{Introduction}
The classical (i.e. non-quantum) equilibrium statistical mechanics 
deals in general with two basic kinds of models: discrete lattice 
systems and continuous fluids.
In one spatial dimension (1D), both kinds of statistical models,
considered with short-range as well as long-range pairwise interactions
among constituents, are solvable in many cases \cite{Lieb}.
In 2D, there is a large family of integrable lattice
systems exactly solvable via the Bethe-ansatz method
(see reviews \cite{Baxter} and \cite{Gaudin}).
On the other hand, there was no exactly solved fluid in more than 1D.
The only partial exceptions were represented by 2D logarithmic models
of Coulomb fluids, the one-component plasma \cite{Jancovici1}
and the symmetric two-component plasma (or Coulomb gas) 
\cite{Cornu} in the point-particle limit, solvable at one 
special value of the dimensionless inverse temperature $\beta=2$.

The situation was changed in very recent years.
In a series of works, the bulk and surface thermodynamics as well as
the large-distance asymptotic behavior of the two-body correlation functions
were derived exactly for the 2D Coulomb gas of pointlike particles,
in the whole stability range of inverse temperatures $\beta<2$.
These results were obtained by mapping the Coulomb gas onto
the 2D sine-Gordon theory with a conformal normalization of
the cos-field, and subsequently applying techniques and recent
achievements in that integrable theory.

The aim of this short review is to present the exact solution 
of the 2D Coulomb gas to the fluid community, in the language
of fluid physics and in a way accessible to non-specialists
in field theory.
Unsolved problems and further potential developments are pointed out.

In section 2, we introduce the 2D Coulomb gas.
Section 3 deals with its complete bulk thermodynamics,
obtained from the mapping onto the bulk sine-Gordon 
field theory.
Surface thermodynamic properties of a semi-infinite 
2D Coulomb gas in contact with an impermeable
(ideal dielectric or ideal conductor) wall, and the
corresponding mapping onto integrable boundary
sine-Gordon models are given in section 4.
The large-distance asymptotic behavior of two-body
correlation functions in the 2D Coulomb gas
is presented in section 5.
Section 6 is devoted to miscellaneous topics
and perspectives. 

\section{Basic facts about the 2D Coulomb gas}
The symmetric Coulomb gas, defined in an infinite 2D space
of points $\br \in R^2$, consists of point particles $\{ i \}$
of charge $\{ q_i = \pm 1 \}$ immersed in a homogeneous
medium of dielectric constant $=1$.
The interaction energy of particles is
$\sum_{i<j} q_i q_j v(\vert \br_i-\br_j \vert)$,
where the Coulomb potential $v$ is the solution of
the 2D Poisson equation
\begin{equation} \label{2.1}
\Delta v(\br) = - 2\pi \delta(\br) .
\end{equation}
Explicitly, $v(\br) = - \ln(\vert \br \vert / L)$ where $L$
is a length scale.
This definition of the Coulomb potential in 2D maintains many
generic properties (e.g. sum rules) of "real" 3D Coulomb fluids with 
the interaction potential $v(\br) = 1/\vert \br \vert$, $\br\in R^3$.

The model is treated as the classical one in thermodynamic
equilibrium, via the grand canonical ensemble characterized
by the (dimensionless) inverse temperature $\beta$ and by
the couple of particle fugacities $z_+(\br)=z_-(\br)=z$.
We set the free length scale $L$ to unity for simplicity;
the true dimension of the rescaled $z$ is then 
[length]$^{\beta/2-2}$.
The grand partition function is defined by
\begin{equation} \label{2.2}
\fl
\Xi = \sum_{N_+,N_-=0}^{\infty} \frac{1}{N_+! N_-!}
\int \prod_{i=1}^N \left[ \rmd^2 r_i~ z_{q_i}(\br_i) \right]
\exp \Big[ -\beta \sum_{i<j} q_i q_j v(\vert \br_i-\br_j \vert)
\Big]
\end{equation}
where $N_+$ $(N_-)$ is the number of positive (negative)
particles and $N=N_+ + N_-$.
Many-particle densities are generated from $\Xi$ in a standard
way as functional derivatives with respect to the fugacity field
$z_q(\br)$, taken at the constant $z_+(\br) = z_-(\br) = z$.
At one-particle level, one introduces the number density of 
particles of one sign 
$n_q = \langle \sum_i \delta_{q,q_i} \delta(\br - \br_i) \rangle$.
Due to the charge symmetry, $n_+ = n_- = n/2$ ($n$ is the total
density of particles).
At two-particle level, one introduces the two-body densities
$n_{qq'}(\vert \br-\br' \vert) = \langle \sum_{i\ne j}
\delta_{q,q_i} \delta(\br-\br_i) \delta_{q',q_j} 
\delta(\br'-\br_j) \rangle$.
It is useful to consider also the pair distribution functions
$g_{qq'}(r) = n_{qq'}(r)/(n_q n_{q'})$, the (truncated) correlation
functions $h_{qq'}(r) = g_{qq'}(r)-1$ and the Ursell functions
$U_{qq'}(r) = n_q n_{q'} h_{qq'}(r)$.

The underlying system of pointlike particles is stable against
the collapse of positive-negative pairs of charges provided that 
the corresponding Boltzmann factor $r^{-\beta}$ is integrable
at short distances in 2D, i.e. for $\beta<2$.
To cross the collapse point $\beta=2$, the pure Coulomb interaction
has to be regularized by a short-distance repulsion, e.g. a hard-core
potential of diameter $\sigma$ around each particle
(the particular choice of the short-distance regularization
influences the results quantitatively, but not qualitatively).
For small values of the dimensionless density $n\sigma^2$, the system
remains in its conducting phase (an external charge is perfectly
screened by the system charges) up to the Kosterlitz-Thouless (KT)
transition of infinite order \cite{Kosterlitz} at a specific
density-dependent $\beta_{\rm KT}$; $\beta_{\rm KT}=4$ in the
low-density limit.
In the insulating phase $\beta>\beta_{\rm KT}$, the system charges
form dipoles and no longer screen an external charge.
At high enough density, the KT critical line splits into a
first order liquid-gas coexistence curve \cite{Levin}.
In what follows, we shall restrict ourselves to the point-particle
Coulomb gas in the stability region $\beta<2$.

A complete exact analysis can be done in two cases: 
in the high-temperature Debye-H\"uckel
limit $\beta \to 0$, and just at the collapse point
$\beta=2$ \cite{Cornu} which corresponds to the free-fermion
point of an equivalent 2D Thirring model.
Although, at a given $z$, the free energy diverges, Ursell functions 
are finite at $\beta=2$.

As concerns an exact information valid in the whole stability
region $\beta<2$, through a simple scaling argument, the exact
equation of state for the pressure $P$,
$\beta P = n ( 1 - \beta/4)$,
has been known for a long time \cite{Salzberg}.
While the density derivatives of the Helmholtz free energy,
like the pressure, can all be calculated exactly, the temperature
derivatives, like the internal energy or the specific heat,
are nontrivial quantities.
Their evaluation can be based on an explicit density-fugacity
$(n,z)$ relationship.
The latter was constructed systematically around the $\beta\to 0$
point by using a bond-renormalized Mayer expansion in density
\cite{Samaj1}: the original bonds $-\beta v(r)$ are summed up
in series to produce the renormalized bonds of strength
$-\beta K_0(\kappa r)$, where $K_0$ denotes a modified Bessel function
and $\kappa=(2\pi\beta n)^{1/2}$ is the inverse Debye length.
The cluster integrals converge in the renormalized format.
The first few integrals imply

\begin{equation} \label{2.3}
\fl
\frac{n^{1-\beta/4}}{z} = 2 \beta^{\beta/4}
\exp \left\{ \left[ 2C + \ln \left( \frac{\pi}{2} \right) \right]
\frac{\beta}{4} + 
\frac{7}{6} \zeta(3) \left( \frac{\beta}{4} \right)^3 +
\zeta(3) \left( \frac{\beta}{4} \right)^4 + O(\beta^5) \right\}
\end{equation}
where $C$ is the Euler number and $\zeta$ denotes the Riemann
zeta function.

\section{Bulk thermodynamics}
The 2D Coulomb gas is equivalent to the 2D sine-Gordon model
\cite{Minnhagen}.
Introducing the microscopic charge density
${\hat \rho}(\br) = \sum_{i=1}^N q_i \delta(\br - \br_i)$,
the interaction energy can be written as
\begin{equation} \label{3.1}
E = \frac{1}{2} \int \rmd^2 r \rmd^2 r' {\hat \rho}(\br)
v(\vert \br-\br' \vert) {\hat \rho}(\br') - \frac{1}{2} N v(0) .
\end{equation}
Let us forget for a while that $v(0)$ diverges, and renormalize
the fugacity by the self-energy term $\exp[-\beta v(0)/2]$,
without changing the $z$-notation.
Using the fact that $-\Delta/(2\pi)$ is the inverse operator of
$v(\br)$ [see equation (\ref{2.1})], the grand partition function
(\ref{2.2}) with $z_q(\br)=z$ can be turned via the 
Hubbard-Stratonovich transformation into
\numparts
\begin{equation} \label{3.2a}
\Xi(z) = \frac{\int {\cal D}\phi \exp\left( - S(z) \right)}{\int
{\cal D}\phi \exp\left( - S(0) \right)}
\end{equation}
where
\begin{equation} \label{3.2b}
S(z) = \int \rmd^2 r \left[ \frac{1}{16\pi} 
\left( \nabla \phi \right)^2 - 2 z \cos (b\phi) \right] ,
\qquad b = \left( \frac{\beta}{4} \right)^{1/2}
\end{equation}
\endnumparts
is the Euclidean action of the 2D sine-Gordon theory.
Here, $\phi(\br)$ is a real scalar field and $\int {\cal D}\phi$
denotes the functional integration over this field.
The many-particle densities are expressible as the averages
over the sine-Gordon action as follows
\begin{equation} \label{3.3}
n_q = z_q \langle \rme^{\rmi q b \phi} \rangle , \qquad
n_{qq'}(\vert \br-\br' \vert) = z_q z_{q'} \langle 
\rme^{\rmi q b \phi(\br)} \rme^{\rmi q' b \phi(\br')} \rangle
\end{equation}
etc.
The parameter $z$ in (\ref{3.2b}), i.e. the fugacity renormalized
by the diverging self-energy term, gets a precise meaning when one
fixes the normalization of the coupled cos-field.
In the Coulomb system, the behavior of the two-body density for
oppositely charged particles is dominated at short distance by
the Boltzmann factor of the Coulomb potential,
$n_{+-}(\br,\br') \sim z_+ z_- \vert \br-\br' \vert^{-\beta}$
as $\vert \br-\br' \vert \to 0$.
With regard to (\ref{3.3}), the mapping is supplemented by
the short-distance normalization
\begin{equation} \label{3.4}
\langle \rme^{\rmi b \phi(\br)} \rme^{-\rmi b \phi(\br)}
\rangle \sim \vert \br-\br' \vert^{-4 b^2} \qquad {\rm as}\
\vert \br-\br' \vert \to 0
\end{equation}
which was usually omitted in the statmech literature.
Under this short-distance normalization, the divergent self-energy 
factor disappears from statistical relations calculated within 
the sine-Gordon representation.
This can be easily verified in the Debye-H\"uckel limit $\beta\to 0$,
when $\cos(b\phi) \sim 1-b^2\phi^2/2$, and the consequent
Gaussian field theory reproduces the $n,z$ relation (\ref{2.3}) up
to the linear $\beta$-term in the exponential.

In the classical limit of the sine-Gordon theory, only such
configurations of the $\phi$-field are considered which fulfill
the equation of motion $\delta S = 0$.
This classical limit is integrable \cite{Zamolodchikov1}, 
i.e. there exists an infinite sequence of conserved quantities.
Due to the discrete symmetry $\phi \to \phi + 2\pi n/b$
($n$ integer), the model has an infinite number of vacua
at $\phi_n = 2\pi n/b$.
The basic "particles", the soliton $S$ and antisoliton ${\bar S}$ pair
of equal masses $M$, interpolate between two neighboring vacua.
The $S-{\bar S}$ pair can create bound states, called the
breather particles $\{ B \}$.
The sine-Gordon model is integrable at the full "quantum level" 
(all configurations of the $\phi$-field are considered) as well
\cite{Zamolodchikov1}, with the same particle spectrum.
The essential difference between the classical and quantum theories
is that the breathers become quantized, $\{ B_j; j=1,2,\ldots < 1/\xi \}$, 
and their number depends on the inverse of 
the parameter $\xi = b^2/(1-b^2)$.
The mass of the $B_j$-breather is given by
\begin{equation} \label{3.5}
m_j = 2 M \sin \left( \frac{\pi \xi}{2} j \right)
\end{equation}
and this breather disappears from the spectrum just when $m_j = 2 M$.
Note that breathers exist only in the stability region of the Coulomb
gas $0<b^2<1/2$ $(0<\beta<2)$.
The lightest $B_1$-breather disappears just at $b^2=1/2$ $(\beta=2)$,
which is the field-theoretical manifestation of the collapse phenomenon.

Using the Thermodynamic Bethe ansatz, the dimensionless specific
grand potential
\begin{equation} \label{3.6}
\lim_{V\to \infty} \frac{1}{V} \ln \Xi =
\frac{m_1^2}{8\sin(\pi\xi)}
\end{equation}
was found by Destri and de Vega \cite{Destri}.
Under the conformal normalization (\ref{3.4}), the relationship between
the soliton mass $M$ and the fugacity $z$ was established in 
\cite{Zamolodchikov2},
\begin{equation} \label{3.7}
z = \frac{\Gamma(b^2)}{\pi \Gamma(1-b^2)} \left[ M
\frac{\sqrt{\pi} \Gamma\left( (1+\xi)/2\right)}{2\Gamma(\xi/2)}
\right]^{2-2 b^2}
\end{equation}
where $\Gamma$ stands for the Gamma function.
Equations (\ref{3.5})--(\ref{3.7}) constitute a complete set
to be solved for the exact $n,z$ relationship \cite{Samaj1}:
\begin{equation} \label{3.8}
\fl
\frac{n^{1-\beta/4}}{z} = 2 \left( \frac{\pi\beta}{8} \right)^{\beta/4}
\frac{\Gamma(1-\beta/4)}{\Gamma(1+\beta/4)} 
\left[ F\left( \frac{1}{2},\frac{\beta}{4-\beta};
1+\frac{\beta}{2(4-\beta)};1\right) \right]^{1-\beta/4}
\end{equation}
where $F\equiv {_2F_1}$ is the hypergeometric function.
The expansion of the rhs around $\beta\to 0$ reproduces correctly
the first terms of the renormalized expansion (\ref{2.3}).
For fixed $z$, the particle density given by (\ref{3.8}) exhibits
the expected collapse singularity $n \sim 4\pi z^2/(2-\beta)$
as $\beta\to 2^-$.
This behavior can be derived independently by combining an
electroneutrality sum rule,
$-q n_q = \sum_{q'=\pm} q' \int \rmd^2 r n_{qq'}(r)$,
with the short-distance asymptotic behavior of $n_{+-}(r)$ discussed above.
Since the derivation of formula (\ref{3.7}) was based on
special analyticity assumptions, the check of the results from
both sides of the stability interval is important. 
Without noticing it, such checks are made for all presented results.

Based on the explicit $n,z$ relation, one can pass by using 
the Legendre transformation from the grandcanonical to 
the canonical ensemble, to obtain the Helmholtz free energy.
We present the explicit result for the excess specific heat at
constant volume per particle \cite{Samaj1},
\begin{equation} \label{3.9}
\fl
\eqalign{
\frac{c_V^{\rm ex}}{k_{\rm B}} & =  \frac{\beta}{4} 
+ \frac{4}{4-\beta} + \frac{\beta^2}{16} 
\left[ \psi'\left( 1-\frac{\beta}{4} \right) -
\psi'\left( 1+\frac{\beta}{4} \right) \right] \\
& - \frac{2\beta^2}{(4-\beta)^3} 
\left[ \psi'\left( \frac{2}{4-\beta} \right) -
\psi'\left( \frac{8-\beta}{8-2\beta} \right) \right]
- \frac{4\pi^2\beta^2}{(4-\beta)^3}
\frac{\cos\left( \pi\beta/(4-\beta) \right)}{\sin^2\left(
\pi\beta/(4-\beta)\right)} . }
\end{equation}
Here, $\psi(x) = \rmd\left[ \ln \Gamma(x) \right]/\rmd x$
is the psi function and $\psi'(x) = \sum_{i=0}^{\infty} 1/(x+i)^2$.
As $\beta\to 2^-$, the $c_V^{\rm ex}/k_{B}$ exhibits the expected
singularity of type $2/(2-\beta)^2$.
Notice that the specific heat is independent of the particle density,
which is a peculiarity of the 2D Coulomb gases.

\section{Surface thermodynamics}
Let us now consider a semi-infinite 2D Coulomb gas in the Cartesian
half-space $x>0$, in contact with a hard wall of dielectric constant
$\epsilon_W$ in the complementary half-space $x<0$.
The presence of the dielectric wall manifests itself through
particle images \cite{Jackson}: the particle of charge $q$ at position 
$\br = (x,y)$ has the image of charge $q^*$ (dependent on $\epsilon_W$)
at $\br^* = (-x,y)$.
We will consider two particular cases: the ideal dielectric wall 
$(\epsilon_W = 0)$ with image charges $q^*=q$ and the ideal conductor wall 
$(\epsilon_W \to \infty)$ with image charges $q^*=-q$.
Let us introduce the microscopic charge plus image-charge density
${\hat \rho}(\br) = \sum_{i=1}^N q_i \left[ \delta(x-x_i) \pm
\delta(x+x_i) \right] \delta(y-y_i)$; hereinafter, the upper $(+)$
sign corresponds to $\epsilon_W=0$ and the lower $(-)$ sign to
$\epsilon_W\to \infty$.
The interaction energy of the particle-image system can be written 
in both cases as
\begin{equation} \label{4.1}
E = \frac{1}{4} \int \rmd^2 r \int \rmd^2 r'
{\hat \rho}(\br) v(\vert \br-\br' \vert) {\hat \rho}(\br')
- \frac{1}{2} N v(0)
\end{equation}
where the integrations over $\br$ and $\br'$ are taken over 
the whole 2D space.

The form of the interaction energy (\ref{4.1}) resembles 
the one in (\ref{3.1}),
and one can proceed in close analogy with the bulk mapping. 
The grand partition function is expressible as
\numparts
\begin{equation} \label{4.2a}
\Xi(z) = \frac{\int {\cal D}\phi \exp\left( - S(z) \right)}{\int
{\cal D}\phi \exp\left( - S(0) \right)}
\end{equation}
where the $\phi$-field is defined in the whole 2D space and
the nonlocal action reads
\begin{equation} \label{4.2b}
S(z) = \int \rmd^2 r \left[ \frac{1}{16\pi} (\nabla \phi)^2
- 2 z \cos\left( \frac{b}{\sqrt{2}} \left[ \phi(x,y) \pm \phi(-x,y) \right]
\right) \right] 
\end{equation}
\endnumparts
$b=\sqrt{\beta/4}$.
To make this field theory local, we introduce two new fields
\begin{equation} \label{4.3}
\fl
\phi_e(x,y) = \frac{1}{\sqrt{2}} \left[ \phi(x,y) + \phi(-x,y) \right] ,
\qquad \phi_o(x,y) = \frac{1}{\sqrt{2}} \left[ \phi(x,y) - \phi(-x,y) \right]
\end{equation}
defined only in the positive $x\ge 0$ half-space.
The "even" field has a Neumann boundary condition,
$\partial_x \phi_e \vert_{x=0} = 0$, 
and the "odd" field a Dirichlet boundary condition,
$\phi_o \vert_{x=0} = 0$.
Since $\int \rmd^2 r (\nabla \phi)^2 = \int_{x>0}
\rmd^2 r \left[ (\nabla \phi_e)^2 + (\nabla \phi_o)^2 \right]$,
the fields $\phi_e$ and $\phi_o$ are decoupled in the
action (\ref{4.2b}).
The field, contributing only by its free-field part,
disappears from (\ref{4.2a}) by numerator-denominator cancellation.
Consequently, renaming the kept field by $\phi$, we arrive at
\begin{equation} \label{4.4}
\fl
\Xi(z) = \frac{\int {\cal D} \phi \exp\left( -S(z) \right)}{\int
{\cal D} \phi \exp\left( -S(0) \right)} , \qquad
S(z) = \int_{x>0} \rmd^2 r \left[ \frac{1}{16\pi} (\nabla \phi)^2
- 2 z \cos(b\phi)\right] .
\end{equation}
Here, the $\phi$-field has the boundary condition: 
$\phi \vert_{x=0} = 0$ for the ideal conductor wall;
$\partial_x \phi \vert_{x=0} = 0$ for the ideal dielectric wall.
The mapping onto the boundary sine-Gordon model is supplemented
by the short-distance normalization (\ref{3.4}).
Both cases under consideration belong to the integrable 
boundary field theories \cite{Ghoshal}.
The thermodynamic quantity of interest is the surface tension $\gamma$,
which characterizes the surface part of the grand potential
$\Omega = - \beta^{-1} \ln \Xi$.

The problem of the ideal conductor wall
was solved via a lattice regularization of the boundary
sine-Gordon model \cite{Samaj2}, namely the XXZ Heisenberg quantum
chain in boundary magnetic fields.
The surface tension was obtained in terms of the soliton mass $M$
as follows
\begin{equation} \label{4.5}
\beta \gamma_{\rm cond} = \frac{M}{4} \left\{ 1
- \tan\left( \frac{\pi\beta}{2(4-\beta)} \right)
- \left[ \cos\left( \frac{\pi\beta}{2(4-\beta)} \right) \right]^{-1}
\right\} .
\end{equation}
The surface collapse is governed by the interaction Boltzmann factor
of a particle with its self-image, $x^{-\beta/2}$.
The 1D integral $\int \rmd x x^{-\beta/2}$ diverges at short
distances at point $\beta=2$ (which coincides with the bulk collapse
point), and this is indeed the radius of convergence of $\gamma_{\rm cond}$.

The problem of the ideal dielectric wall was solved
by exploring a "reflection" relationship between the Liouville
and sine-Gordon theories \cite{Samaj3}.
The result is
\begin{equation} \label{4.6}
\beta \gamma_{\rm diel} = \frac{M}{4} \left\{ 1
+ \tan\left( \frac{\pi\beta}{2(4-\beta)} \right)
- \left[ \cos\left( \frac{\pi\beta}{2(4-\beta)} \right) \right]^{-1}
\right\} .
\end{equation}
At $\beta=2$, $\gamma_{\rm diel}$ keeps a finite value
\cite{Jancovici2}.
The analytic continuation of (\ref{4.6}) beyond the bulk collapse point
predicts a surface collapse at $\beta=3$.
Such a phenomenon is due to a paradoxical short-distance attraction of 
a particle with its own image charge of the same sign \cite{Samaj3}.

The surface thermodynamics of a plain hard wall $(\epsilon_W=1)$,
which is the last exactly solvable case at the free-fermion point
$\beta=2$ \cite{Cornu2}, is an open problem.

\section{Large-distance behavior of particle correlations}
In a 2D integrable theory characterized by a "particle" spectrum,
correlation functions of local fields can be written as an infinite
convergent series over multi-particle intermediate states, in terms
of the corresponding form-factors.
The form-factor representation is especially useful at large distances,
since the dominant contribution of the series comes from an
intermediate state with the minimum value of the total particle mass.
In the sine-Gordon theory, for topological reasons, solitons $S$ and
antisolitons ${\bar S}$ coexist in pairs, the total mass of the pair
being $2M$.
The breathers $\{ B_j \}$, when they exist, have lighter masses
[see formula (\ref{3.5})].
The lightest $B_1$-breather with mass $m_1=2M\sin(\pi\xi/2)$, which
exists in the whole stability region of the Coulomb gas,
governs the large-distance asymptotic behavior 
of the two-point correlation in (\ref{3.3}).
In particular, one has for the correlation function $h_{qq'}$
\cite{Samaj4}, as $r\to\infty$,
\begin{equation} \label{5.1}
h_{qq'}(r) \sim q q' h(r) , \qquad
h(r) = - \lambda \left( \frac{\pi}{2 m_1 r} \right)^{1/2}
\exp (- m_1 r)
\end{equation}
where $\lambda$ is a $\beta$-dependent prefactor.
The specific dependence of $h_{qq'}(r)$ on the charge product $qq'$
means that the two-particle correlations are determined at large
distance by the charge-charge correlation function
$h_{\rho}(r) = (1/4) \sum_{q,q'=\pm} q q' h_{qq'}(r)$ $= h(r)$.

On the other hand, the density correlation function
$h_n = (1/4) \sum_{q,q'=\pm} h_{qq'}$ vanishes for the lowest
one-$B_1$ breather state, and becomes nonzero only for the next
two-$B_1$ breathers state \cite{Samaj5}.
The mass of two $B_1$-breathers, $2 m_1$, is smaller than the one
of the soliton-antisoliton pair, $2M$, in the region $0<\beta<1$.
Consequently, as $r\to\infty$,
\begin{equation} \label{5.2}
h_n(r) \propto \cases{\exp(-2 m_1 r) & for $0 < \beta < 1$ ; \\
\exp(-2 M r) & for $1 \le \beta < 2$ . \\}
\end{equation}
The correlation length depends continuously on $\beta$, but its
first derivative with respect to $\beta$ is discontinuous at
$\beta=1$.
The large-distance exponential decay of $h_n$ is faster than
the one of $h_{\rho}$.
The two correlation lengths coincide just at the collapse point
$\beta=2$, where $h_{\rho}$ and $h_n$ differ from one another
only by the inverse-power law prefactors.

\section{Miscellaneous topics and perspectives}
The 2D Coulomb gas exhibits, at any temperature in the stability region, 
a universal finite-size correction to the grand potential as if we had 
a critical theory with the conformal anomaly number $c=-1$, although 
the particle correlation functions presented here decay exponentially.
This phenomenon follows intuitively from the sine-Gordon representation 
of $\Xi$ (\ref{3.2a}), with the critical massless Gaussian field theory 
$(c=1)$ in the denominator.
The explicit checks of the critical-like behavior were done at the 
free-fermion $\beta=2$ point for various geometries of confining domains 
\cite{Forrester2,Jancovici3,Jancovici4}, and at any $\beta<2$
for the sphere \cite{Jancovici5,Jancovici6} and for the disk \cite{Samaj5}.

The ultimate task is to solve exactly the 2D Coulomb gas with a
short-distance (maybe temperature-dependent) regularization
of the pure Coulomb potential.
We have made a first step toward this aim by deriving
the leading correction to the exact bulk thermodynamics of
pointlike charges due to presence of the hard core of diameter
$\sigma$ around particles \cite{Kalinay}.
The results, which are conjectured to be exact in the low-density
limit up to $\beta=3$, reproduce correctly the $\sigma$-singularities
of thermodynamic quantities at $\beta=2$.
They also confirm a "subtraction" mechanism of singularities
between the collapse point $\beta=2$ and the KT transition point
$\beta_{\rm KT}=4$ within an ansatz proposed by Fisher et al.
\cite{Fisher} (excluding the existence of an infinite number
of intermediate phases proposed in \cite{Gallavotti}),
however, predict a different analytic structure of this ansatz.
There exist candidates among integrable 1D quantum systems,
for example the lattice sine-Gordon model \cite{Essler}, which,
after being formulated as a 2D Euclidean theory, might represent
a Coulomb gas regularized in the whole temperature range.

Another topic, which attracted much attention in the last years
due to the phenomenon of charge inversion \cite{Shklovskii}, 
is the charge asymmetry. 
The Coulomb gas with $\vert q_+\vert = 2\vert q_-\vert$ was solved
in terms of the equivalent complex Bullough-Dodd field theory
in reference \cite{Samaj6}.
Here, the fundamental changes in statistics caused by the charge asymmetry 
(e.g. the same correlation length for both correlation functions 
$h_{\rho}$ and $h_n$ at any stable $\beta$) were documented.
This result might be a motivation for the exact solution of
the 2D one-component plasma, which is the extreme
charge-asymmetry case of the 2D Coulomb gas.

\ack
I am grateful to Bernard Jancovici for his support and
careful reading of the manuscript.
I thank the Organizing Committee of the SCCS conference for 
the invitation.
A partial financial coverage by Grant VEGA 2/7174/20 is acknowledged.


\section*{References}

\begin{thebibliography}{32}

\bibitem{Lieb} Lieb E H and Mattis D C 1966 {\it Mathematical
Physics in One Dimension} (New York: Academic Press);
Mattis D C 1993 {\it The Many-Body Problem}
(Singapore: World Scientific)

\bibitem{Baxter} Baxter R J 1982 {\it Exactly Solved Models
in Statistical Mechanics} (London: Academic Press) 

\bibitem{Gaudin} Gaudin M 1983 {\it La Fonction d'Onde de Bethe}
(Paris: Masson)

\bibitem{Jancovici1} Jancovici B 1981 
{\it Phys. Rev. Lett.} {\bf 46} 386

\bibitem{Cornu} Cornu F and Jancovici B 1987 
{\it J. Stat. Phys.} {\bf 49} 33

\bibitem{Kosterlitz} Kosterlitz J M and Thouless D J 1973
{\it J. Phys. C} {\bf 6} 1181

\bibitem{Levin} Levin Y, Li X J and Fisher M E 1994
{\it Phys. Rev. Lett.} {\bf 73} 2716

\bibitem{Salzberg} Salzberg A and Prager S 1963
{\it J. Chem. Phys.} {\bf 38} 2587

\bibitem{Samaj1} \v Samaj L and Trav\v enec I 2000
{\it J. Stat. Phys.} {\bf 101} 713

\bibitem{Minnhagen} Minnhagen P 1987
{\it Rev. Mod. Phys.} {\bf 59} 1001

\bibitem{Zamolodchikov1} Zamolodchikov A B and Zamolodchikov Al B 1979
{\it Ann. Phys. (N.Y.)} {\bf 120} 253

\bibitem{Destri} Destri C and de Vega H 1991
{\it Nucl. Phys. B} {\bf 358} 251

\bibitem{Zamolodchikov2} Zamolodchikov Al B 1995
{\it Int. J. Mod. Phys. A} {\bf 10} 1125

\bibitem{Jackson} Jackson J D 1998
{\it Classical Electrodynamics}, 3rd ed. (New York: Wiley)

\bibitem{Ghoshal} Ghoshal S and Zamolodchikov A B 1994
{\it Int. J. Mod. Phys. A} {\bf 9} 3841

\bibitem{Samaj2} {\v S}amaj L and Jancovici B 2001
{\it J. Stat. Phys.} {\bf 103} 717

\bibitem{Samaj3} {\v S}amaj L 2001
{\it J. Stat. Phys.} {\bf 103} 737

\bibitem{Jancovici2} Jancovici B and {\v S}amaj L 2001
{\it J. Stat. Phys.} {\bf 104} 755

\bibitem{Cornu2} Cornu F and Jancovici B 1989
{\it J. Chem. Phys.} {\bf 90} 2444

\bibitem{Samaj4} {\v S}amaj L and Jancovici B 2002
{\it J. Stat. Phys.} {\bf 106} 301

\bibitem{Samaj5} {\v S}amaj L and Jancovici B 2002
{\it J. Stat. Phys.} {\bf 106} 323

\bibitem{Forrester2} Forrester P J 1991
{\it J. Stat. Phys.} {\bf 63} 491

\bibitem{Jancovici3} Jancovici B, Manificat G and Pisani C 1994
{\it J. Stat. Phys.} {\bf 76} 307

\bibitem{Jancovici4} Jancovici B and T\'ellez G 1996
{\it J. Stat. Phys.} {\bf 82} 609

\bibitem{Jancovici5} Jancovici B 2000
{\it J. Stat. Phys.} {\bf 100} 201

\bibitem{Jancovici6} Jancovici B, Kalinay P and {\v S}amaj L 2000
{\it Physica A} {\bf 279} 260

\bibitem{Kalinay} Kalinay P and {\v S}amaj L 2002
{\it J. Stat. Phys.} {\bf 106} 857

\bibitem{Fisher} Fisher M E, Li X J and Levin Y 1995
{\it J. Stat. Phys.} {\bf 79} 1

\bibitem{Gallavotti} Gallavotti G and Nicol\'o F 1985
{\it J. Stat. Phys.} {\bf 39} 133

\bibitem{Essler} Essler F H L, Frahm H, Its A R and Korepin V E 1997
{\it J. Phys. A: Math. Gen.} {\bf 30} 219

\bibitem{Shklovskii} Shklovskii B I 1999
{\it Phys. Rev. E} {\bf 60} 5802

\bibitem{Samaj6} {\v S}amaj L 2003
{\it J. Stat. Phys.} {\bf 111} 261

\end{thebibliography}
\end{document}